\documentclass[12pt]{article}%
\usepackage[nosort]{cite}
\usepackage{graphicx}
\usepackage{multicol}
\usepackage{amsfonts}
\usepackage{amssymb}
\usepackage{amsmath}
\usepackage{heck}
\usepackage{afterpage}
\usepackage{setspace}
\usepackage{verbatim}
\usepackage{color}
\usepackage{longtable}
\usepackage{float}
\usepackage{subcaption}
\usepackage{epsfig}
\usepackage{epstopdf}
\usepackage{adjustbox}
\usepackage{tikz}
\usepackage[margin=1in]{geometry}
\usepackage{titletoc}
\usepackage{hyperref}%
\providecommand{\U}[1]{\protect\rule{.1in}{.1in}}
\pdfoutput=1
\newsavebox{\mysavebox}

\hypersetup{colorlinks,citecolor=black,filecolor=black,linkcolor=black,urlcolor=black,pdftex}
\usetikzlibrary{decorations.markings}

\numberwithin{equation}{section}

\newcommand{\ba}{\begin{eqnarray}}
\newcommand{\ea}{\end{eqnarray}}

\newcommand{\be}{\begin{equation}}
\newcommand{\ee}{\end{equation}}
\tikzstyle{startstop} = [rectangle, rounded corners, minimum width=3cm, minimum height=1cm,text centered, draw=black, fill=blue!10]
\tikzstyle{startstop} = [rectangle, rounded corners, minimum width=3cm, minimum height=1cm,text centered, draw=black, fill=blue!10]
\tikzstyle{io} = [trapezium, trapezium left angle=70, trapezium right angle=110, minimum width=3cm, minimum height=1cm, text centered, draw=black, fill=blue!30]
\tikzstyle{process} = [rectangle, minimum width=3cm, minimum height=1cm, text centered, draw=black, fill=orange!30]
\tikzstyle{decision} = [diamond, minimum width=3cm, minimum height=1cm, text centered, draw=black, fill=green!30]
\tikzstyle{arrow} = [thick,->,>=stealth]
\begin{document}

\date{December 2015}

\title{$750$ GeV Diphotons from a D3-brane}

\institution{UNC}{\centerline{${}^{1}$Department of Physics, University of North Carolina, Chapel Hill, NC 27599, USA}}

\institution{COLUMBIA}{\centerline{${}^{2}$Department of Physics, Columbia University, New York, NY 10027, USA}}

\institution{CUNY}{\centerline{${}^{3}$CUNY Graduate Center, Initiative for the Theoretical Sciences, New York, NY 10016, USA}}

\authors{Jonathan J. Heckman\worksat{\UNC, \COLUMBIA, \CUNY}\footnote{e-mail: {\tt jheckman@email.unc.edu}}}

\abstract{Motivated by the recently reported diphoton excess at $750$ GeV observed by both CMS and ATLAS, we study string-based
particle physics models which can accommodate this signal. Quite remarkably, although
Grand Unified Theories in F-theory tend to impose tight restrictions on candidate extra sectors, the case of a probe D3-brane near an
E-type Yukawa point naturally leads to a class of strongly coupled models capable of accommodating the observed signature.
In these models, the visible sector is realized by intersecting 7-branes, and the $750$ GeV resonance is a scalar modulus associated
with motion of the D3-brane in the direction transverse to the Standard Model 7-branes.
Integrating out heavy $3-7$ string messenger states leads to dimension five operators for
gluon fusion production and diphoton decays. Due to the unified structure of interactions, these models
also predict that there should be additional decay channels to $ZZ$ and $Z \gamma$. We also comment on models with distorted unification,
where both the production mechanism and decay channels can differ.}

\maketitle

\enlargethispage{\baselineskip}

\setcounter{tocdepth}{2}

\newpage

\section{Introduction \label{sec:INTRO}}

Recently, the LHC\ experiments CMS\ and ATLAS have both announced tentative evidence for a diphoton excess with a
resonant mass near $750$ GeV \cite{CMS:2015, ATLAS:2015}. This signal is seen in the early data of the $13$ TeV run, and appears to be
compatible with the absence of a large signal in the earlier $7$ and $8$ TeV runs.
Recall that the observed diphoton signal depends on the production cross section $\sigma_{pp\rightarrow s}$
for the resonance ``$s$,'' as well as $B_{s\rightarrow\gamma\gamma}=\Gamma_{s\rightarrow\gamma\gamma
}/\Gamma_{s\rightarrow\text{any}}$ its branching fraction to diphotons:
\begin{equation}
\sigma_{pp\rightarrow s}\times B_{s\rightarrow\gamma\gamma}\sim5\text{ fb}.
\end{equation}
In the case of the ATLAS\ experiment, there is
also an even more preliminary indication that this resonance may have a
substantial width.

While the observed signal is on the order of three sigma (if one naively
combines CMS\ and ATLAS), this is still far from meeting the threshold for
discovery. Even so, it has already inspired a number of theoretical analyses (see e.g.
\cite{Cai:2015bss, Harigaya:2015ezk, Mambrini:2015wyu, Backovic:2015fnp, Nakai:2015ptz, Knapen:2015dap, Buttazzo:2015txu,
Pilaftsis:2015ycr, Franceschini:2015kwy, DiChiara:2015vdm, Higaki:2015jag, McDermott:2015sck, Ellis:2015oso, Low:2015qep, Bellazzini:2015nxw,
Gupta:2015zzs, Petersson:2015mkr, Molinaro:2015cwg, Dutta:2015wqh, Cao:2015pto, Matsuzaki:2015che, Kobakhidze:2015ldh, Martinez:2015kmn,
Cox:2015ckc, Becirevic:2015fmu, No:2015bsn, Demidov:2015zqn, Chao:2015ttq, Fichet:2015vvy,
Curtin:2015jcv, Bian:2015kjt, Chakrabortty:2015hff, Ahmed:2015uqt, Agrawal:2015dbf,
Csaki:2015vek, Falkowski:2015swt, Aloni:2015mxa, Bai:2015nbs, Gabrielli:2015,
Benbrik:2015, Kim:2015, Alves:2015, Megias:2015, Carpenter:2015, Han:2015qqj, Murphy:2015kag}).
One of the lessons which can already be drawn
from these early phenomenological studies is that in general (but not always),
models with some strongly coupled extra sector appear
to fare better in generating a sufficiently large signal with a broader decay width.
From this perspective, it is natural to ask whether there are UV motivated
constructions of new physics which can accommodate the diphoton excess.

In this note we point out that string-based constructions from F-theory Grand
Unified Theories (F-theory GUTs) naturally suggest particular strongly coupled
extra sectors which can easily accommodate the diphoton excess. We stress
that this is non-trivial, since the underlying exceptional gauge symmetries necessary for stringy
unification tightly constrain both the structure of the
visible sector matter content, as well as possible extra sectors. Indeed,
experience from earlier constructions such as reference \cite{Heckman:2009mn}
shows that intersecting 7-branes can realize the visible sector, but
little else. Rather, extra probe D3-branes must typically be included to get novel
phenomenology from an extra sector \cite{Heckman:2010fh}. The resulting physics is
quite rich, and leads to several novel features. First, these models
are strongly coupled, but nevertheless, preserve supersymmetric gauge coupling unification
\cite{Heckman:2011hu}. Additionally, depending on the mass
scales available, they can lead to rather striking phenomenological signatures. One
of our aims will be to show how this class of models can naturally interpolate
between several of the simplified models presented which have been used to
explain the diphoton excess.

\section{Extra Sector from a D3-Brane}

In more detail, we consider models of particle physics which embed in a
supersymmetric Grand Unified Theory in F-theory known as an ``F-theory GUT'' \cite{Beasley:2008dc, Donagi:2008ca}. For a
review of the relevant particle physics constructions, see for example
\cite{Heckman:2010bq}. Though order TeV scale supersymmetry is not essential for most
of our discussion, it is well motivated. It will also make details of the model more calculable, so we
shall assume approximate supersymmetry in the extra sector as a convenient computational tool.

\begin{figure}[ptb]%
\centering
\includegraphics[scale = 0.5, trim={0 3cm 0 3cm}]{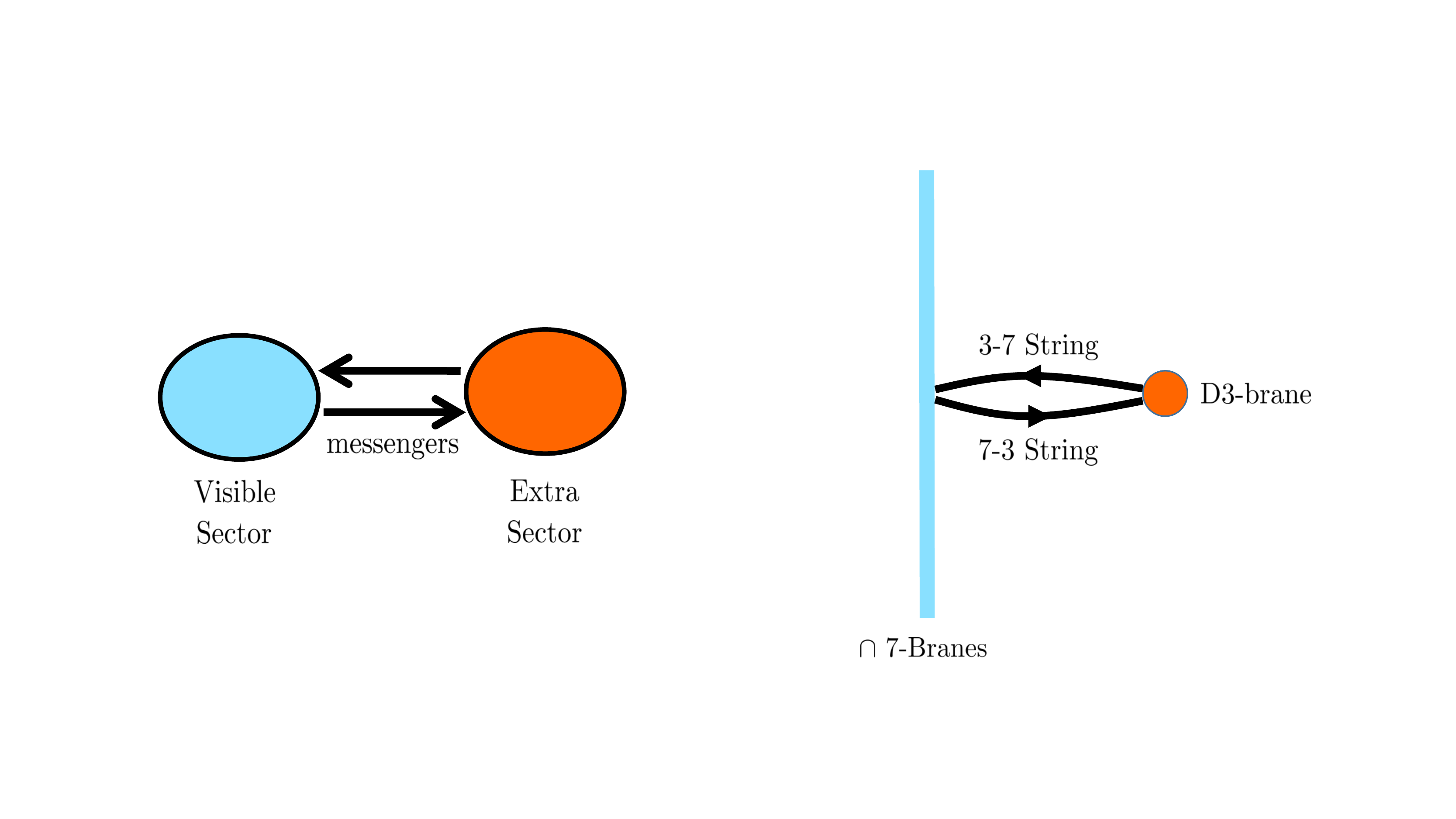}%
\caption{Depiction of how strongly coupled extra sectors arise in F-theory
GUTs. The Standard Model comes from intersecting 7-branes, and a probe
D3-brane provides an extra sector.\ The two communicate with each other via
$3-7$ strings, i.e. \textquotedblleft messenger states.\textquotedblright}%
\label{messengers}%
\end{figure}

In these models, the visible sector is realized on a stack of intersecting
7-branes (i.e. spacetime filling branes which fill four extra dimensions).
The extra sector is given by a D3-brane (i.e. a spacetime filling brane which
is pointlike in the extra dimensions) probing the Standard Model (SM) stack. The
same mechanism which generates subleading Yukawa couplings for the SM also
generates a potential for D3-branes with a local minimum near the Yukawa point of the
SM stack \cite{Heckman:2010fh}. See figure \ref{messengers} for a depiction.

%EndExpansion

In F-theory, grand unification requires unbroken exceptional gauge symmetries
at subspaces of the internal dimensions, which in turn
demands that the string coupling is order one. So, the extra sector on
the D3-brane is always strongly coupled. If the D3-brane is at a generic point
of the SM\ stack, then we get a $U(1)$ gauge theory, and
messengers transforming as a supersymmetric vectorlike generation
in the $\mathbf{5}\oplus\overline{\mathbf{5}}$ of an $SU(5)$ GUT.

If the D3-brane localizes at the special points of unbroken exceptional symmetry (as expected from
the mechanism used to generate flavor physics in the model),
additional light states enter the spectrum, and we instead get a strongly
coupled conformal field theory \cite{Heckman:2011hu, Heckman:2010qv}. This is
given by an $\mathcal{N} = 1$ deformation of an $\mathcal{N} = 2$ superconformal
field theory (SCFT) with $E_8$ flavor symmetry known as the ``Minahan-Nemeschansky theory'' \cite{Minahan:1996fg, Minahan:1996cj}.
The Standard Model gauge group arises from weakly gauging an $SU(3) \times SU(2) \times U(1)$ subgroup of $E_8$.
There can also be order one couplings between the Higgs sector and the third (i.e. heaviest) generation of SM states.

For the purposes of explaining the diphoton excess, our interest in this class
of models is the generic feature that they are strongly coupled (i.e. $g^{2}_{\mathrm{extra}} / 4 \pi \sim O(1)$),
and that there are states which are charged under both the SM\ gauge group, and the extra sector
$U(1)_{\text{extra}}$ of the D3-brane which we
refer to as $3-7$ strings or ``messenger states.''\footnote{In string theory,
these messengers actually arise from multi-prong strongly coupled bound states
of fundamental strings and D1-branes. Nevertheless, we shall find it helpful
to use this concise characterization.} To emphasize that these strings couple to the SM
gauge groups, we shall sometimes write $3-7_{\mathrm{vis}}$.

An additional important feature is that although they are strongly coupled, some
still preserve supersymmetric gauge coupling unification at the percent level \cite{Heckman:2011hu}.
This is perhaps not too surprising when the messengers are very heavy, i.e. $10^{13}$ GeV,
since the overall magnitude of the threshold correction is suppressed by a moderate sized logarithm. In the case
where the messengers are far lighter, i.e. $1$ TeV, this log-running is more pronounced, but remarkably enough
there are \textit{still} a few models where precision unification is still respected at the percent level.
As a benchmark model of this type, we shall often focus on the case of the ``$Dih_{4}^{(2)}$ monodromy model'' of reference \cite{Heckman:2011hu}.\footnote{The name of the model has to do with the details of how the visible sector is constructed,
i.e. it is the Galois group of the spectral equation for a matrix valued position dependent complex scalar which controls the profile of intersecting 7-branes. Other choices such as the $S_3$ and $Dih_{4}^{(1)}$ monodromy models lead to order ten percent threshold corrections when running from the TeV to GUT scale, which is still tolerable considering there could be additional thresholds at both the TeV and GUT scale.} For this model, the effects of the heavy messenger states contribute a threshold correction on the order of $2.2$
supersymmetric vector-like generations in the $\mathbf{5} \oplus \overline{\mathbf{5}}$ of $SU(5)_{GUT}$.

This extra sector includes various mass
scales which can lead to phenomenologically relevant effects. Of particular
significance is the complex scalar $S$ which controls the position of the
D3-brane in the direction transverse to the 7-brane.
Activating a non-zero vev $f \equiv \sqrt{2} \left\langle S\right\rangle\neq0$, all
of these $3-7$ states pick up a mass of rough order $4 \pi f$, so depending on the overall value of
this vev, these states could be near the TeV scale, or far higher, i.e. near
the GUT scale \cite{Heckman:2011hu}, with the details depending on the structure of
the potential for the D3-brane. There is a general expectation that if
supersymmetry breaking occurs in the visible sector in the $1-10$ TeV range (as would be expected
in a model of approximate low energy supersymmetry), then the mass scale $4 \pi f$ will
also be on the order of the TeV scale. On general grounds, we also expect the colored states
to be somewhat heavier than their color-neutral counterparts, simply due to SM loop corrections.

There are also SM neutral states from both $3-7_{\mathrm{flav}}$ and $3-3$ strings. Here, a $7_{\mathrm{flav}}$-brane
is one which acts as an approximate flavor symmetry for the F-theory GUT model, which is supported
on a $7_{\mathrm{vis}}$-brane. Indeed, the intersection between a $7_{\mathrm{flav}}$-brane and a
$7_{\mathrm{vis}}$-brane leads to localized matter of the Standard Model (i.e. the quarks and leptons). Now,
when the $U(1)_{\mathrm{extra}}$ gauge symmetry is unbroken,
a seesaw  mechanism tends to makes some of the $3-7_{\mathrm{flav}}$ and $3-3$ strings much lighter than the $3-7_{\mathrm{vis}}$
states \cite{Heckman:2011sw}. In addition to this scalar vev, there are also electric and
magnetically charged states of $U(1)_{\text{extra}}$. Supersymmetry
breaking usually causes this $U(1)$ to be broken, and so
depending on the details of this process, the extra photon could either be light
(i.e. below $750$ GeV / $2$ so that it is a candidate decay mode), or could be quite heavy (i.e. above the TeV scale).

Finally, there is also the mass of $S$ itself. As already mentioned, we expect
that at least near the GUT scale, we can approximate the dynamics of the
D3-brane by an $\mathcal{N}=1$ SCFT with a Coulomb branch scalar $\widetilde{S}$.
Non-perturbative instanton corrections
can generate a superpotential for this modulus, which has the leading order
behavior:%
\begin{equation}
W(\widetilde{S})=m\widetilde{S}^{2}+...
\end{equation}
Even though this superpotential deformation is
quadratic in $\widetilde{S}$, it is actually an irrelevant deformation of the SCFT. The
reason is that $\widetilde{S}$ will typically have dimension $\Delta$ greater than $3/2$. Indeed,
for the benchmark $Dih_{4}^{(2)}$ model mentioned above, we have $\Delta \sim 2$.
Working in terms of a canonically normalized field
$S=(M_{\text{GUT}})^{\Delta-1} \times \widetilde{S}$, we find that the effective mass of the
excitation is of order:%
\begin{equation}
m_{S}\sim M_{\text{GUT}}\cdot\left(  \frac{M_{\text{IR}}}{M_{\text{GUT}}%
}\right)  ^{\Delta-1},
\end{equation}
i.e. we run down to the scale of conformal symmetry breaking and calculate the size of the
perturbation in the infrared. So, for $\Delta \sim 2$, a value of $M_{\mathrm{IR}} \sim 1$ TeV produces a TeV
scale mass for $S$. Depending on how supersymmetry is broken in the extra sector, the two real degrees of freedom in $S$
can have different masses.

Summarizing, we see that in this class of models, there are generically a few
different characteristic mass scales, which we summarize as $M_{3-7_{\mathrm{vis}}}$, $M_{3-7_{\mathrm{flav}}}$,
$M_{3-3}$, $M_{U(1)}$, $M_{S}$. Depending on how we adjust these parameters,
we can expect various types of phenomenological scenarios. For illustrative purposes, we shall
consider first the case where we have the most analytic control, i.e. where the various mass scales are all heavier than that
of $S$. In other limits, we still generate a diphoton excess though we have less quantitative control over the model.

\subsection{Minimal D3-Brane Models}

To give an example, consider the special case where we can approximate the
effects of the $3-7$ strings as much heavier, i.e. $M_{3-7}\gg M_{S}$. In this
case, we can treat these charged states as giving a threshold correction, and
we integrate them out of the low energy theory. Since the mass of the
threshold is controlled by $\left\langle S\right\rangle $, we can also read off the
dimension five operator which couples $S$ to the SM\ gauge fields.
In the holomorphic approximation of references \cite{Heckman:2012nt, Heckman:2012jm},
i.e. when the effects of wave function renormalization are small (as is indeed the case for us) we get:
\begin{equation}\label{integrateout}
L\supset\underset{G}{\sum}\operatorname{Re}\int d^{2}\theta\text{ }%
\frac{\delta b_{G}}{32\pi^{2}}\log S\text{ Tr}\mathcal{W}_G^{\alpha
}\mathcal{W}^G_{\alpha},
\end{equation}
where here all gauge algebra generators are embedded in the standard way in $SU(5)_{GUT}$, and
the sum runs over the three simple gauge factors of the SM. The
overall size of $\delta b_{G}$ depends on the specific D3-brane probe theory,
but in our benchmark $Dih_{4}^{(2)}$ model, $\delta b_{G}\sim 2.2$.
In our normalization, this amounts to roughly $2.2$ supersymmetric vector-like generations
in the $\mathbf{5}\oplus\overline{\mathbf{5}}$ of $SU(5)_{GUT}$. There are also subleading
order $1 \%$ differences between the threshold corrections.

Let us now turn to the expected couplings of $S$ with the SM\ gauge fields.
The same methods developed in references \cite{Heckman:2012nt, Heckman:2012jm, Heckman:2011bb}
(for earlier work see \cite{Shifman:1979eb}) to explore possible enhancements of couplings to the
Higgs sector are readily adapted.\footnote{We note that in general, the size of the cutoff means that there may also be interesting
corrections to Higgs sector couplings. It would be interesting to explore these signatures further in future work.}
Expanding out in terms of:
\begin{equation}
S= \frac{1}{\sqrt{2}}(f+s+ia),
\end{equation}
we get the interaction terms:\footnote{We note that including an arbitrary scale $\mu$
in the logarithm of line (\ref{integrateout}) i.e. $\log(S/\mu)$
does not change the resulting couplings.}
\begin{equation}
L\supset\underset{G}{\sum}-\frac{1}{2g_{G}^{2}}\text{Tr}F_G^{\mu\nu}%
F^G_{\mu\nu}+\frac{\delta b_{G}}{32\pi^{2}}\left(  \frac{s}{f}\right)  \text{
Tr}F_G^{\mu\nu}F^G_{\mu\nu}+\frac{\delta b_{G}}{32\pi^{2}}\left(  \frac{a}%
{f}\right)  \text{ Tr}F_G^{\mu\nu}\widetilde{F}^G_{\mu\nu}.
\end{equation}
By inspection, we see that there are couplings to all of the SM\ gauge
fields for both the real and imaginary parts of $S$. In particular, we expect
there to be production and decay channels with strength set by the overall
size of the threshold correction. Some examples important for current and
upcoming experiment are:

\begin{itemize}
\item $pp\rightarrow s/a\rightarrow\gamma\gamma$

\item $pp\rightarrow s/a\rightarrow gg$

\item $pp\rightarrow s/a\rightarrow ZZ$

\item $pp\rightarrow s/a\rightarrow WW$

\item $pp\rightarrow s/a\rightarrow Z\gamma$
\end{itemize}
where the decay rates will be primarily set by the ratios of the gauge
couplings -- at least for a multiplet structure with approximate gauge
coupling unification. As alluded to above, based on the structure of the
model, we see that there can even be two nearly degenerate resonances (with $s$
CP-even and $a$ CP-odd), though this can of course
be split by supersymmetry breaking effects.

When $s$ is the lightest mode of $S$, the phenomenology is actually quite
close to the \textquotedblleft hidden glueball model\textquotedblright\ and
when $a$ is the lightest mode of $S$, we instead get the \textquotedblleft
hidden pion model\textquotedblright\ of reference \cite{Harigaya:2015ezk}. The primary
features common to both our models is a strongly coupled extra sector with
multiplets compatible with gauge coupling unification. Additionally, in the
D3-brane context, we can expect some additional decay channels to light SM neutral states such as
$3-3$ states and $3-7_{\mathrm{flav}}$ states. These masses are in turn dependent on the breaking
scale for the extra $U(1)$. Depending on whether the extra $U(1)$ is light enough, there can also be
additional decay channels to SM neutral extra sector states. This
can lead to a significant enhancement in the width of $S$ but also a
decrease in the branching fraction $B_{s \rightarrow \gamma \gamma}$.

Working in the computationally most tractable regime where we decouple these additional extra states, there are no
decays to hidden sector states and we can simply match the
parameters of our model to that of reference \cite{Harigaya:2015ezk}. Doing so, we get that
the production cross section times the branching fraction to diphotons is roughly:%
\begin{equation}
\sigma_{pp\rightarrow s}B_{s\rightarrow\gamma\gamma}\sim 1.3 \text{ fb}%
\times\left(  \delta b_{\text{GUT}}\right)  ^{2}\times\left(  \frac{1\text{
TeV}}{4\pi f}\right)  ^{2}.
\end{equation}
In the benchmark ``$Dih_4^{(2)}$ model'' $\delta b_{G}\sim 2.2$, so to match to
an observed excess of order $5$ fb, we need to set $4\pi f\sim 1.1$ TeV,
which is not altogether surprising considering that the mass of the
resonance is $750$ GeV. Even so, we expect the threshold approximation
adopted here to be valid due to the fact that the leading order
wave function renormalization effects have already been taken into account in the
quantity $\delta b_{\text{GUT}}$. For other probe D3-brane theories with a larger threshold
correction (i.e. $\delta b_{G} \sim 3 $), the value of $4 \pi f$ is somewhat higher, though there is then a bit more
tension with precision unification considerations.\footnote{For example, in the $S_3$ 7-brane monodromy model, $\delta b_{G} \sim 3$ and
we get $4 \pi f \sim 1.5$ TeV, while for the $Dih_4^{(1)}$ 7-brane monodromy model, $\delta b_{G} \sim 2.4$ and $4 \pi f \sim 1.2$ TeV. In these cases, however, there is a somewhat bigger threshold correction to precision unification of order $9 \%$ to $12 \%$, respectively.}
In any case, this would indeed suggest that exciting the $3-7$ string
states may be within reach in the near term.

Another general comment is that even when we work in this decoupled limit, the
overall decay rate to gluons is going to be bigger than in the case of a
weakly coupled messenger model. Roughly speaking, it is as if the messengers
had \textquotedblleft non-perturbative Yukawas\textquotedblright\ to the field
$S$. Fitting to the examples of weakly coupled messengers presented in
reference \cite{Knapen:2015dap}, we see that the overall width in this regime is on the
order of $10-100$ MeV. This is compatible with overall $pp \rightarrow s/a \rightarrow jj$ limits, see e.g.
\cite{Franceschini:2015kwy}. For earlier discussion of constraints from production via gluon fusion and
decay to diphotons, see e.g. \cite{Jaeckel:2012yz}.

Now, as we lower the value of $4 \pi f$, the overall mass scale for the
$3-7$ strings will become lighter. When we do this, the threshold
approximation adopted above will start to break down, and indeed, there can even be cascade decays from a
heavy messenger state to $S$. The signature space for this class of models has
recently appeared in reference \cite{Knapen:2015dap}, to which we refer the interested reader.
We can also consider the case where we decrease the $U(1)_{\mathrm{extra}}$ breaking scale.
When we do this, it is also expected that some of the SM neutral states of the extra
sector will also become quite light \cite{Heckman:2011sw}, and we can expect qualitatively similar phenomenology to the case
of a hidden valley scenario (see e.g. \cite{Strassler:2006im}) of the type considered in \cite{Knapen:2015dap}.

Finally, another important feature of this class of models is that because the states
organize according to supersymmetric GUT multiplets, we should also expect a decay rate to
$s \rightarrow ZZ\rightarrow4\ell$ and $s \rightarrow Z\gamma \rightarrow 2 \ell \gamma$ though
presently, current limits do not impose much of a constraint. This appears to be a common feature of many
of the earlier phenomenological models considered in \cite{Cai:2015bss, Harigaya:2015ezk, Mambrini:2015wyu, Backovic:2015fnp, Nakai:2015ptz, Knapen:2015dap, Buttazzo:2015txu, Pilaftsis:2015ycr, Franceschini:2015kwy, DiChiara:2015vdm, Higaki:2015jag, McDermott:2015sck, Ellis:2015oso, Low:2015qep, Bellazzini:2015nxw, Gupta:2015zzs, Petersson:2015mkr, Molinaro:2015cwg}.

\subsection{Models with Distorted Unification}

Though the models based on supersymmetric unification are more elegant (and
also far easier to construct), it is also of interest to
consider how far we can distort this structure to accommodate possible phenomenological signatures. Indeed,
depending on whether the reported large width from ATLAS is confirmed, and if an eventual signal is seen in other
channels, this would give a way to further narrow the list of options presented in this note.

First of all, there are effects coming from various threshold corrections, i.e. how we generate the various dimension five operators
for production and decay. In most of the unified models, this appears to be dominated by gluon fusion production, but as pointed out
for example in \cite{Fichet:2015vvy, Csaki:2015vek, Aloni:2015mxa},
this may make it difficult to accommodate the preliminary indication of a
$45$ GeV width state, a point we can confirm at least in the models we have studied. Though we lose analytic control over various aspects of the model, one could envision that in some extreme region of the mass scales of the D3-brane probe theory, there is a sufficient mass splitting in the various thresholds so that the low energy physics is dominated by the coupling to the $U(1)_Y$ gauge boson, in which case photon fusion may become the dominant production mechanism \cite{Fichet:2015vvy, Csaki:2015vek}.
In such models, a strongly coupled extra sector is still quite helpful,
a feature which is manifestly present in the D3-brane construction.

One can also contemplate more extreme distortions where one simply
works with a stack of intersecting 7-branes with no apparent
unification at high scales. For concreteness, suppose that we have at least
three stacks of branes, as in the quiver model of reference \cite{Berenstein:2006pk}.
Then, by moving the D3-brane close to the stacks where say $SU(3)$ and
a $U(1)$ factor are localized, we can raise the suppression scale of the dimension five operator which couples $S$ to
$SU(2)$ gauge bosons. This would in turn suppress the decay modes $s \rightarrow ZZ$ and $s \rightarrow Z \gamma$ since
now they must proceed through the $U(1)_Y$ gauge boson. Alternatively, we can move the D3-brane to regions where it only touches the
$U(1)$ stack, which would provide a way to generate the signature primarily through photon fusion.

\section{Conclusions}

The recent hints of a $750$ GeV diphoton excess at CMS\ and ATLAS is quite exciting.
There are by now many proposed models which aim to accommodate this excess. Here, we have
taken a different approach, asking whether considerations from strings can guide us to a particular class of
motivated choices. In this note we have pointed
out that in F-theory GUTs, there are some preferred classes of models.
These models are based on introducing an additional
probe D3-brane close to the stack of intersecting 7-branes used to
engineer the Standard Model. Integrating out messenger states yields dimension five
operators of precisely the kind needed to explain the diphoton excess.
We have also seen that various distortions in unification
can lead to some deviations from this simplest class of models, though
the qualitative feature of a strongly coupled extra sector remains.
We find it encouraging that rather than positing
an \textquotedblleft ad hoc\textquotedblright\ extra structure, the necessary
ingredients to explain the diphoton excess are already a part of many F-theory GUT models.

\section*{Acknowledgements}

We thank C. Kilic, D. Pappadopulo, J. Ruderman and N. Weiner for helpful discussions.
JJH also thanks the theory groups at Columbia University, the ITS at the CUNY\ graduate center, and the CCPP at
NYU for hospitality during the completion of this work. The work of JJH is supported by NSF CAREER grant PHY-1452037.
JJH also acknowledges support from the Bahnson Fund at UNC Chapel Hill.

%%%%%%%%%%%%%%%%%%%%%%%%%%%%%%%%%%%%%%%%%%%%%%%%%%%%%%%%%%%%%%%%%%%%%%%%%%%%

\bibliographystyle{utphys}
\bibliography{D3photons}

\end{document}